\documentclass[10pt,journal,compsoc]{IEEEtran}

\ifCLASSOPTIONcompsoc
\usepackage[nocompress]{cite}
\else
\usepackage{cite}

\fi

\ifCLASSINFOpdf
\else
\fi

\usepackage{listings}
\usepackage[]{xcolor}
\usepackage{color,soul}
\usepackage{hyperref}
\usepackage{soul}
\usepackage[font=small,skip=1.5pt]{caption}
\usepackage{notoccite}
\usepackage{filecontents,lipsum}
\usepackage{graphicx}
\usepackage{listings}
\usepackage{amsmath}
\usepackage{color}
\definecolor{lightgray}{rgb}{.9,.9,.9}
\definecolor{darkgray}{rgb}{.4,.4,.4}
\definecolor{purple}{rgb}{0.65, 0.12, 0.82}

\usepackage{subfigure}

\usepackage[english]{babel}

\usepackage{booktabs}

\usepackage{colortbl}

\usepackage{comment}

\usepackage{graphicx}

\usepackage{epsfig}

\usepackage{array, colortbl}

\usepackage{listings}

\usepackage{epstopdf}

\usepackage{multirow}

\usepackage{rotating}

\usepackage{xspace}

\usepackage{float}

\usepackage{amsmath}

\usepackage{balance}

\usepackage{fancybox}

\usepackage{scalefnt}

\usepackage{amssymb}

\usepackage[normalem]{ulem}

\usepackage{mathtools}

\usepackage{enumitem}

\usepackage{tikz}

\usepackage{tabularx,multirow,booktabs,blindtext}

\usepackage{graphicx}

\usepackage{mdframed}

\usepackage{lipsum}

\usepackage[utf8]{inputenc}

\usepackage{caption}

\usepackage{footnote}

\usepackage[roman]{parnotes}

\usepackage{array}

\usepackage{cleveref}

\usepackage{url}

\newcolumntype{C}[1]{>{\centering\arraybackslash}p{#1}}

\makeatletter

\renewcommand{\paragraph}[1]{\noindent\textsf{#1}.}

\definecolor{britishracinggreen}{rgb}{0.0, 0.26, 0.15}

\definecolor{cadmiumorange}{rgb}{0.93, 0.53, 0.18}

\definecolor{darkbyzantium}{rgb}{0.36, 0.22, 0.33}

\definecolor{darkgreen}{rgb}{0.0, 0.2, 0.13}

\definecolor{oxfordblue}{rgb}{0.0, 0.13, 0.28}

\definecolor{palatinateblue}{rgb}{0.15, 0.23, 0.89}

\definecolor{pansypurple}{rgb}{0.47, 0.09, 0.29}

\newcommand*{\rom}[1]{\expandafter\@slowromancap\romannumeral #1@}

\newcommand{\npm}{{npm}\xspace}

\newcommand\rqone{RQ$_1$: How are the threat levels of vulnerable dependencies distributed in the studied Node.js applications?
}
\newcommand\rqtwo{RQ$_2$: How does the distribution of the threat levels change as the studied applications evolve?}

\newcommand\rqthree{RQ$_3$: Who is responsible for the dependence on high threat vulnerable dependencies? 
}
\begin{document}
	%
	\title{	\LARGE On the Threat  of \npm Vulnerable Dependencies in Node.js Applications}
	%
	%
	%
	%
	
	\author{Mahmoud~Alfadel,
		Diego~Elias~Costa, 
		Mouafak~Mokhallalati,
		Emad~Shihab,~\IEEEmembership{Senior Member,~IEEE} and Bram~Adams,~\IEEEmembership{Senior Member,~IEEE}

		\IEEEcompsocitemizethanks{\IEEEcompsocthanksitem Mahmoud Alfadel,  Diego Elias Costa, Mouafak Mukhallalati, and Emad Shihab are with the Data-driven Analysis of Software (DAS) Lab at the Department of Computer Science and Software Engineering, Concordia University, Montr\'{e}al, Canada.
			
			
			\IEEEcompsocthanksitem Bram Adams is with the Lab on Maintenance, Construction, and Intelligence of Software (MCIS), D\'{e}partement de G\'{e}nie Informatique et G\'{e}nie Logiciel, \'{E}cole Polytechnique de Montr\`{e}al, Montr\'{e}al, Canada.
			
		}
		
		\thanks{Manuscript received xxx; revised xxx.}}

	\IEEEtitleabstractindextext{%
		\begin{abstract}
			Software vulnerabilities have a large negative impact on the software systems that we depend on daily. Reports on software vulnerabilities always paint a grim picture, with some reports showing that 83\% of organizations depend on vulnerable software. However, our experience leads us to believe that, in the grand scheme of things, these software vulnerabilities may have less impact than what is reported.

			Therefore, we perform a study to better understand the threat of npm vulnerable packages used in Node.js applications. We define three threat levels for vulnerabilities in packages, based on their lifecycle, where a package vulnerability is assigned a \textit{low} threat level if it was hidden or still unknown at the time it was used in the dependent application (\textit{t}), \textit{medium} threat level if the vulnerability was reported but not yet published at \textit{t}, and \textit{high} if it was publicly announced at \textit{t}.
			Then, we perform an empirical study involving 6,673 real-world, active, and mature open source Node.js applications.
			Our findings show that although 67.93\% of the examined applications depend on at least one vulnerable package, 94.91\%  of the vulnerable packages in those affected applications are classified as having low threat. Moreover, we find that in the case of vulnerable packages classified as having high threat, it is the application's lack of updating that makes them vulnerable, i.e., it is not the existence of the vulnerability that is the real problem. Furthermore, we verify our findings at different stages of the application's lifetime and find that our findings still hold. Our study argues that when it comes to software vulnerabilities, things may not be as bad as they seem and that considering vulnerability threat is key. 
		\end{abstract}
		
		}

	\maketitle

	\IEEEdisplaynontitleabstractindextext

	%
	\IEEEpeerreviewmaketitle

	\section{Introduction}
	\label{sec:introduction}
The existence of a software vulnerability in a software system is a major concern for software projects. These vulnerabilities can cause unimaginable damage for an organization if exploited. In fact, there are many examples of such cases. 
One such example is the Equifax cybersecurity incident~\cite{EquifaxR95}, where a vulnerability in Apache Struts led to unauthorized access to consumers' personal information and credit card numbers.

To make matters even worse, the recent popularity of software ecosystems has only magnified the problem. Specifically, most software systems today have many direct and transitive dependencies, which increases the risk of a vulnerability in a software project. Contrast Security, a software security company, reported that 80\% of the code written in today's applications depend on external packages, and approximately one fourth of package downloads have known vulnerabilities~\cite{williams2012unfortunate}. Furthermore, a recent report by Snyk.io showed that 83\% of organizations use vulnerable packages and that 77\% of the 430,000 websites crawled by them, run at least one vulnerable JavaScript package~\cite{77of433093:online}.
These reported figures are worrisome given our everyday dependence on software systems.
 
However, although these vulnerability reports are worrying, they impact a very tiny fraction of existing software systems~\cite{pashchenko2018vulnerable,zapata2018towards}.
For example, a recent study manually analysed 60 projects that depend on high severity vulnerabilities, and found that 73.3\% of them were actually safe because they didn't make use of the vulnerable functionality of their dependencies~\cite{zapata2018towards}.
%

Hence, we argue that \emph{not all vulnerabilities are equal}. To get the real picture, one needs to take into consideration the \emph{potential threat} of a software vulnerability. Formally defined, the threat of a vulnerability is {the potential danger to exploit a vulnerability in order to breach security and cause possible harm}~\cite{shirey2000internet}.



The main goal of our study is to examine the degree that applications rely on vulnerable dependencies and understand how threatening such vulnerable dependencies really are. To achieve our goal, we first provide a threat classification for the software vulnerabilities based on their lifecycle. Note that this is a post-mortem classification, using information only available after the fact, for the purpose of evaluating the threat of dependency vulnerabilities in the dependent applications. We classify software vulnerabilities into three main threat levels: \emph{low threat}, indicating that a vulnerability that affects a dependency was not discovered (reported) yet at a specific point in the application lifecycle; \emph{medium threat}, indicating that a vulnerability was discovered but not yet published (publicly announced); \emph{high threat}, indicating a vulnerability has been published. 

We use our classification and perform an empirical study involving 6,673 real-world, active, and mature open source Node.js applications, of which more than half have at least one vulnerable dependency. We use these classifications to examine {(RQ1) how threatening vulnerable dependencies in the dependent applications really are,} (RQ2) how the threat levels of vulnerable dependencies evolve through the applications development history, and (RQ3) why some applications end up depending on high threat vulnerabilities in order to better understand how we can mitigate such issues. 

Our findings show that although 67.93\% of the examined applications have (in one of their recent versions) at least one vulnerable dependency, 94.91\% of the vulnerable dependencies in these applications are classified as having low threat (RQ1). Moreover, as applications evolve, they are more likely to depend on vulnerable dependencies, however, most of the vulnerabilities have a low threat level (RQ2). Lastly, we find that the vast majority (90.8\%) of the high threat dependency vulnerabilities were caused by the applications, i.e., vulnerable dependencies had an available vulnerability fix but the applications did not update to a newer (safer) version of the vulnerable dependency (RQ3).





As a key contribution, we provide an empirically-sound evidence regarding the degree to which Node.js application projects rely on npm vulnerable dependencies and how such vulnerable dependencies are threatening through applications development history, while also discussing the implications of our findings to researchers and practitioners. Besides, we provide an approach to identify vulnerable dependencies in a Node.js application at a given point in time, taking into consideration our vulnerability threat classification. Other researchers analysing vulnerabilities in npm dependencies can reuse it. Finally, we provide a replication package comprising the techniques and dataset that we used in this study as a means to bootstrap other studies in the area.

The rest of the paper is organized as follows. Section~\ref{sec:background} introduces our vulnerability classification used in this study. Section~\ref{sec:dependency} describes how npm manages dependencies in Node.js applications. Section~\ref{sec:case_study_design} describes our case study design. Section~\ref{sec:results} presents our results. Section~\ref{sec:discussion} discusses how our results lead to direct implications to researchers and practitioners. Section~\ref{sec:related_work} discusses the related work. Section~\ref{sec:threats} presents the threats to validity. Section~\ref{sec:conclusion} concludes our paper.

	
	\begin{figure*}[!t]
		\centering
	\end{figure*}
	
	\section{Classifying Vulnerabilities}
	\label{sec:background}



In this section, we explain the different stages a vulnerability goes through in its lifecycle. Then, we define our threat levels for vulnerabilities using the different stages that a vulnerability goes through during its lifecycle.

\subsection{Vulnerability  Lifecycle}
\label{sec:background2}
\begin{figure}[tb!]
	\centering
	\includegraphics[width=1\linewidth]{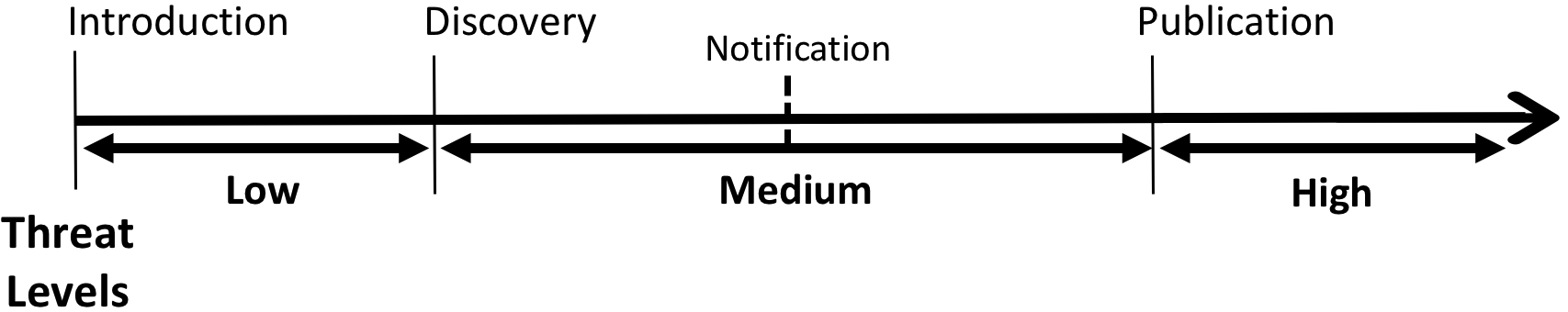}
	\caption[]{Classification of threats over the vulnerability lifecycle.} 
	\label{vuln_timeline}
\end{figure}

A software \emph{vulnerability} is a weakness that allows unauthorized actions and/or access to be performed. These actions are typically used to break through the system and violate its security policies~\cite{liu2012software,shirey2000internet}. A \emph{vulnerability threat} is a 
potential danger to exploit a vulnerability in order to breach security and cause possible harm~\cite{shirey2000internet}.
As shown in Figure~\ref{vuln_timeline}, typically, and with emphasis on vulnerabilities in the Node Package Manager (\npm ecosystem), a vulnerability goes through a number of different stages~\cite{Reporting96}.

\begin{itemize}
\item \textbf{Introduction.} This is when the software vulnerability is first introduced into the code. At this stage, no one really knows about its existence, assuming that the introduction is not malicious. Hence, the potential threat of the vulnerability is quite low. 

\item \textbf{Discovery (report).} When a vulnerability is discovered, it must be reported to the npm security team. The npm team investigates to ensure that the reported vulnerability is legitimate. At this stage, only the security team and the reporter of the vulnerability know about its existence. The  potential threat at this stage is still low.
%
\item \textbf{Notification.} Once the reported vulnerability is confirmed, the security team triages the vulnerability and notifies the vulnerable package maintainers. At this stage, only the reporter, npm team, and package maintainers know about the vulnerability, hence its potential threat to be exploited remains low.

%
%
\item \textbf{Publication without a known fix.} Once the package maintainers are notified, they have 45 days before npm publishes the vulnerability publicly. 
Alongside with publishing the vulnerability, the npm team may also publish a proof-of-concept showing how the vulnerability can be exploited. 
At this stage, the vulnerability is known publicly and its potential threat is high.

\item \textbf{Publication with a fix.} Another (and more common) way that a vulnerability can be published is when a fix is provided by the package maintainers. If a fix is provided (before 45 days), then npm publishes the vulnerability along with the version of the package that fixes the vulnerability. At this stage, the potential threat is not as high as when a no fix is provided, but now the onus is on the application maintainers to make sure that they pull in the latest fixes, otherwise they are risking being exploited.
\end{itemize}

Typically, the vulnerability publish date is after the report and notification dates. It is important to note that although the aforementioned stages are generally sequential, we do see cases where {it is not}. For example, in some cases we see vulnerabilities with a fix date that precedes its reporting or publication
date. The race between developers and attackers starts as soon as a vulnerability is discovered. We use the different stages of a vulnerability to examine the potential threat of software vulnerabilities next.

\subsection{Threat Levels}
\label{sec:ex}

As shown earlier, the different stages that a vulnerability goes through significantly impact its threat. Hence, our study is based on the idea that vulnerabilities should be examined while taking their threat into consideration as the \textit{vulnerability timing} makes them hard to exploit. We use the various stages to ground our argument and define three specific threat levels:

\begin{enumerate}

\item \textbf{Low threat - before discovery (report).} Since very little (or nothing at all) is known about a vulnerability before it is found, i.e., vulnerabilities are hidden in the applications, we believe that its potential threats and chances of being exploited are very low. Hence, we classify all vulnerabilities at this stage as having \emph{low threat}.

\item \textbf{Medium threat - after discovery \& before publication:} Once a vulnerability has been discovered, there is potential that others may also know about it. Moreover, since at this stage the public is still not aware of the vulnerability, 
the vulnerability might be exploited by people who know about it somehow and have the capability to exploit it. Hence, we classify vulnerabilities at this stage as having \emph{medium threat}.

\item \textbf{High threat - after publication:} After publication 
this is the time when the chance of exploitability is highest. Of course, if a fix is provided, then the risk is lower, however, if the application does not update then it still faces a major risk of  being exploited. If a fix is not provided, then all applications are at a very high risk of being exploited, hence, we classify all vulnerabilities at this stage as having \emph{high threat}.
\end{enumerate}

	\section{NPM Dependency Management}
	\label{sec:dependency}
We use our defined threat levels to examine vulnerable dependencies in Node.js applications. Since determining vulnerable dependencies heavily relies on the management of the dependencies and how they are resolved, in this section, we highlight how npm dependency management works.

Node Packages Manager (npm) is the main package manager used by Node.js applications to manage their dependencies~\cite{npmregis47:online}. npm has a registry where packages are published and maintained.
To date, npm registry hosts more than 1.3M packages, and has had the highest growth rate in terms of packages amongst all known programming languages~\cite{npmLibra8}.

To determine the threat of vulnerable dependencies in Node.js applications, we need to 
understand two important mechanisms of the npm ecosystem: 1) how Node.js applications specify their npm dependencies and 2) how npm resolves a dependency version, i.e., find the dependency version to install in a Node.js application.
Node.js applications specify their dependencies in a JSON-format file, called package.json, which lists the dependencies and their versioning constraints. The versioning constraint is a convention to specify the dependency version(s) of the package that an application is willing to depend upon.
The version constraints can be static, requiring a specific version of the dependency (e.g., ``P:1.0.0'' ), or dynamic specifying a range of versions of the dependency (e.g., ``P:$>$1.0.0''). Typically, developers use dynamic versioning constraints if they want to install the latest version of a dependency, allowing them to get the latest updates/security fixes of the package. When a dynamic version is used, the resolved version (i.e., the actual version) corresponds to the latest installable version that satisfies the constraint~\cite{cogo2019empirical}.



Node.js applications can specify two sets of dependencies in their package.json file: development and production dependencies. Development dependencies are installed only on development environments, and consequently, issues that may arise from them (e.g., vulnerabilities and bugs) have no impact on production environments. On the other hand, production dependencies (also called runtime dependencies) are installed on both production and development environments. In our work, we only consider production 
dependencies in our analysis since they are the ones that impact the production environment~\cite{decan2018empirical}.

	
	\section{Case Study Design}
	\label{sec:case_study_design}
	To examine the degree to which applications rely on vulnerable dependencies and how threatening such vulnerable dependencies are within the applications, we study a large dataset of mature and active Node.js applications that use external dependencies.
First, we describe our data collection in Section~\ref{subsec:data}.
Then, in Section~\ref{subsec:scanning} we explain how we use our threat levels to identify and classify the vulnerable dependencies in the Node.js applications. We leverage the collected data to answer the following research questions.
\begin{itemize}
	\item  \rqone 
	\item  \rqtwo 
	\item  \rqthree
\end{itemize}

\subsection{Data Collection}
\label{subsec:data}

Our study examines vulnerable dependencies in Node.js applications. We chose to focus on Javascript due to its wide popularity amongst the development community~\cite{StackOve70:online}. 
\\


\noindent\textbf{Packages vs. Applications.} The software community classifies JavaScript projects into two categories:
\textit{1) packages}, also referred to as libraries, which are included in other applications using dependency management tools to help facilitate and speed up development. 
Packages are referred to as "dependencies" of an application. 
\textit{2) applications} are standalone software projects, which are distinct from libraries, where they are not distributed via a package manager and are typically applications for clients and end users rather than components to build upon.
As mentioned before, the Node.js applications mainly state the packages they depend on (i.e., dependencies) in a file called package.json. 

To perform our study, we leverage two datasets: (1) Node.js applications that use {npm} to manage their dependencies, and (2) Security vulnerabilities that affect npm packages. To do so, we \textbf{(i)} obtain the Node.js applications from GitHub, \textbf{(ii)} extract their dependencies, and \textbf{(iii)} obtain the security vulnerabilities for npm packages from {npm} advisories~\cite{npmadvisories}. The dataset collection took place during May and June of 2019.\\

\noindent\textbf{(i) Applications Dataset.} To analyse a large number of open source JavaScript applications that depend on npm packages and obtain insights on their security vulnerabilities, we mine the GHTorrent dataset~\cite{GHtorrent} and extract information about all Node.js applications hosted on GitHub. The GHTorrent dataset contains a total of 7,863,361 JavaScript projects hosted on GitHub, of which 2,289,130 use npm as their package management platform (i.e., these projects contain a file called package.json). Moreover, since both Node.js \textit{packages} and \textit{applications} can use GitHub as their development repository, and our applications dataset should only contain Node.js \emph{applications}, we filter out the GitHub projects  
that are actually npm \emph{packages} by checking their GitHub URL on the {{npm}} registry. The main reason that we focused on applications and not packages is because packages become exploitable only when used and deployed in an application, i.e., packages do not reside on their own in production, they should be part of applications that make use of them. This filtering excludes 328,343 projects from our list of GitHub projects as they are identified as packages and not applications.

As shown in previous studies~\cite{kalliamvakou2014promises,kula2018developers}, some projects on GitHub are immature, hence, to make this study more reliable we refined the dataset using additional filtering criteria to eliminate such immature projects. In particular, we gather applications that satisfy all the following criteria:

%


\begin{itemize}
\item Non-forked applications, as we do not want to have duplicated project history to bias our analysis.

\item Applications that depend upon more than two dependencies.

\item Applications that have at least 100 commits by more than two contributors, which indicates a minimal level of commit activity.

\item Applications that have had their creation date (first commit) before January 1st 2017.
Since vulnerabilities take on median 3 years to be discovered~\cite{decan2018impact}, applications in our dataset need to have a development history long enough to have had a chance for their vulnerabilities to be discovered.

\item Applications that have had their latest commit after January 1st 2017, as we want to analyze applications that had some level of development in the last 3 years. 

\end{itemize}

After applying these refinement criteria, we end up with 6,673 Node.js applications that make use of npm packages. Table~\ref{statsApplications} shows descriptive statistics on the selected Node.js applications in our dataset. Overall, the applications in our dataset
have a rich development history (a median of 213 commits made by 4 developers and 1,657
days of development lifespan) and make ample use of external
dependencies (a median of 11 dependencies).\\


\noindent\textbf{(ii) Application Dependencies.}
After obtaining the applications dataset, we want to extract the history of dependency changes of all applications.
This is necessary to identify the exact dependency versions that would be installed by the application at any specific point-in-time.
As mentioned in Section~\ref{sec:dependency}, Node.js applications specify their dependencies in the package.json file, which contains 
the dependency list, containing the dependent upon packages and their respective version constraints.
Hence, we extract all changes that touched the package.json file and associate each commit hash and commit date to their respective package.json dependency list, creating a history of dependency changes for all applications.
Note that these dependencies are not yet resolved, that is, we only have the version constraints (not the versions) for the dependencies of each application.
\\

\begin{table}[tb!] 
	\centering 
	\caption{Statistics of the 6,673  studied Node.js applications. 
	}
	\label{statsApplications}
	\begin{tabular}{l|r|r|r|r}
		\toprule
		\textbf{Metric} & \textbf{Min.} & \textbf{Median($\bar{x}$)} & \textbf{Mean($\mu$)} & \textbf{Max.} \\ \midrule
		\textbf{Commits} & 100 & 213 & 384.60 & 53,872 \\
		\textbf{Dependencies} & 3 & 11 & 14.93 & 114 \\
		\textbf{Developers} & 3 & 4 & 5.33 & 52 \\
		\textbf{Lifespan (in days)} & 151 & 1,657 & 1,730.07 & 3,575 \\
		\bottomrule
	\end{tabular}
\end{table}

\noindent\textbf{(iii) NPM Advisories Dataset.} 
To identify Node.js applications that depend on vulnerable packages, we need to collect information on npm vulnerable packages. We resort to the \textit{NPM advisories} registry to obtain the required information about all npm vulnerable packages~\cite{npmadvisories}.
The npm advisories dataset is the official registry for npm vulnerability reports, which contains a number of JavaScript vulnerabilities, specific to the Node.js-platform packages.

This dataset provides several kinds of information about vulnerable packages relevant to our study.
Each report has the affected package name, the package versions affected by the vulnerability, and the versions in which the vulnerabiliy was fixed (safe versions). 
The report also contains both the vulnerability discovered (reported) time and published time, which we use in our approach for identifying and classifying vulnerabilities (Section~\ref{subsec:scanning}).
Note that a vulnerable package could be affected by several vulnerabilities (i.e., a vulnerable package appears with different vulnerability reports due to different vulnerability types).

Our initial dataset contains 654 security reports that cover 601 vulnerable packages. Following the criteria filtration process applied by Decan et al.~\cite{decan2018impact},  we removed 12 vulnerable packages of the type "Malicious Package", because they do not actually introduce vulnerable code. These vulnerabilities are packages with names close to popular packages (a.k.a. typo-squatting) in an attempt to deceive users at installing harmful packages. The 12 vulnerable packages account for 12 vulnerability reports. At the end of this filtering process, we are left with 642 security vulnerabilities reports affecting 589 distinct vulnerable packages. 
These packages have combined 26,462 distinct package versions of which 13,868 are affected by vulnerabilities from our report. 
Table~\ref{dataset} shows the summary statistics for vulnerability reports on npm packages.
 
\begin{table}[tb!]
	\centering
	\caption{Descriptive statistics on the npm advisories dataset.}
	\label{dataset}
	\begin{tabular}{l|r}
		\toprule
          Vulnerability reports & 642 \\
		Vulnerable packages & 589  \\
		Versions of vulnerable packages & 26,462    \\
		Affected versions by vulnerability & 13,868 \\ 
		\bottomrule
	\end{tabular}%
	\vspace{-2mm}
\end{table}

\begin{figure*}[tb!]
	\centering
	\setlength{\abovecaptionskip}{5pt}
	\includegraphics[width=0.7\linewidth]{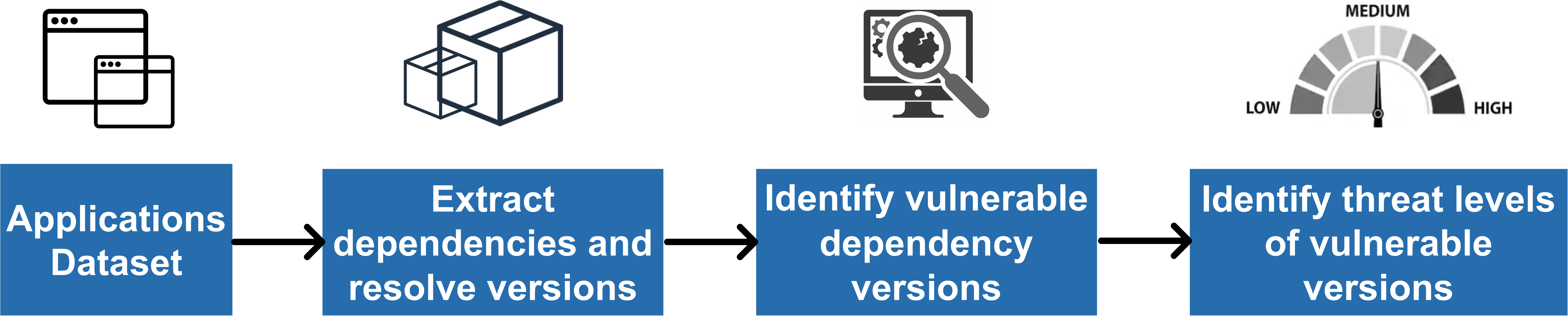}
	\caption[]{Approach for identifying and classifying vulnerable dependencies in Node.js applications.}

	\label{classifyApproach}
\end{figure*}

\subsection{Identifying and Classifying Vulnerable Dependencies in Node.js Applications}
\label{subsec:scanning}
To classify the threat level of vulnerable dependencies at a specific point in the development history of a Node.js application, which we refer to as the \textit{analyzed snapshot time}, we leverage 3-step approach. Figure~\ref{classifyApproach} provides an overview of our approach, which we detail below:



\noindent
{\textbf{Step 1. Extract dependencies and resolve versions.}}
The goal of this step is to extract applications dependencies and find the actual dependency version installed at the analyzed snapshot time.
For each application, we extract the dependency list (with the versioning constraints) at that snapshot time from the history of dependency changes.
After that, to find the actual version of each dependency at the analyzed snapshot,  we utilize the \textit{semver} tool~\cite{semvernp91} that is used by npm to find the latest version that satisfies the versioning constraint, with an additional restriction that the satisfying version should have been released (in the npm registry) before the application snapshot time. For example, an application can specify a versioning constraint (``P:$>$1.0.0'') at the snapshot May 2016. Hence, the actual installed version is the latest version that is greater than 1.0.0 and also has been released in the npm registry before May 2016.
This step allows us to find the installed version of the dependency at the analyzed snapshot time.
\\

\noindent
{\textbf{Step 2. Identify vulnerable dependency versions.}}
After determining the resolved (and presumably installed) version at the analyzed snapshot time, we check whether the resolved version is vulnerably or not. To do so, we check the advisories dataset for the versions that were available at that snapshot point.
If the resolved version is covered by the advisories dataset, we label it as a vulnerable dependency version. We skip the whole next step if the dependency version has not been mentioned in any advisory, i.e., the dependency version is not vulnerable.
\\

\noindent
{\textbf{Step 3. Identify threat levels of vulnerable versions.}}
Once we identify the vulnerable dependency versions at the analyzed snapshot time, we classify each vulnerable dependency version using one of the threat levels we defined earlier (in Section~\ref{sec:ex}), i.e., we find out the threat level of each vulnerable dependency version. To do so, for each vulnerable version, we compare its vulnerability \textit{discovery (report)} and \textit{publication} time to the analyzed snapshot time. As we stated previously (in Section~\ref{sec:ex}), if the vulnerability publication time of the vulnerable dependency version is before the application's snapshot time then we mark the vulnerability as high threat vulnerability. If the vulnerability of the dependency was not published but only discovered (reported) before the application's snapshot time, then we mark it as medium. And finally, if it was neither published nor discovered (reported) before the analyzed snapshot time (i.e., no one knows about it at that snapshot time), then we mark it as low. 

In cases where more than one vulnerability affects the vulnerable dependency version, we resort to a weakest link approach (i.e., we label the vulnerable dependency version with the highest threat level). For example, if we find that the vulnerable version of the dependency is affected by two vulnerabilities -one having low threat and another as high threat, we label the vulnerable dependency version as high at that snapshot time.

\subsection{Replication Package}
To facilitate verification and advancement of research in the field, a replication package comprising the data used in our study along with the analyses used in our study is publicly available\footnote{http://doi.org/10.5281/zenodo.3837397}.
	
	\section{Case Study Results}
	\label{sec:results}
	In this section, we present our case study results that answer our 3 research questions (RQ). For each RQ, we motivate the question, detail the approach used and present the results.



\subsection*{\rqone}
\label{sec:RQ1}
{\textbf{Motivation}}:
Prior work showed that a significant amount of application code comes from third party packages, and a non-negligible amount of these packages are affected by known security vulnerabilities~\cite{williams2012unfortunate}. However, we argue that not all vulnerabilities should be treated equally. Hence, in this RQ we would like to quantify how many of our studied applications have at least one vulnerable dependency and what the threat level of these vulnerable dependencies is. Answering this question will help us understand the real risk/threat of vulnerable packages on the software applications.

%
%
%
\noindent
{\textbf{Approach}}: 
In order to perform an unbiased analysis, we need to account for vulnerability discovery time. Prior work showed that vulnerabilities in npm take on median 3 years to be discovered and publicly announced~\cite{decan2018impact}. As a consequence, selecting snapshots of our applications in 2019 will paint an incomplete picture, as most vulnerabilities recently introduced in the package's code would remain hidden for a median of 3 years. 

Since we collected the advisories dataset in May/June 2019, we chose to evaluate our applications as of May 2016 (3 years prior), which ensures that at least half the dependency vulnerabilities introduced in the code are reported in the current advisories dataset. 

Then, we answer our RQ in two steps. First, we examine if the \textit{selected snapshot} of the application had at least one dependency that contains a vulnerability (irrespective of its threat level). 
Then, to determine the threat level of the vulnerable dependencies in the examined applications, we focus only on the set of applications that have at least one vulnerable dependency using the methodology described in Section~\ref{subsec:scanning}. 
In the second step, we quantify the number of vulnerable dependencies in the applications under each threat level. 
We first check the percentage of overall vulnerable dependencies in each application and illustrate their distribution using a Boxplot. We further analyze the distribution of these vulnerable dependencies across the threat levels and plot it using three Boxplots, one for each threat level. For example, an application could have 10\% of its dependencies as vulnerable at the analysed snapshot, and such percentage (i.e., 10\%) could be distributed across the threat levels as follows: 25\% of the vulnerable dependencies are classified as low threat, 60\% of them are classified as medium, and 15\% as high.

\begin{figure}[tb!]
	\centering
	\includegraphics[width=1\linewidth,height=.85\linewidth]
	{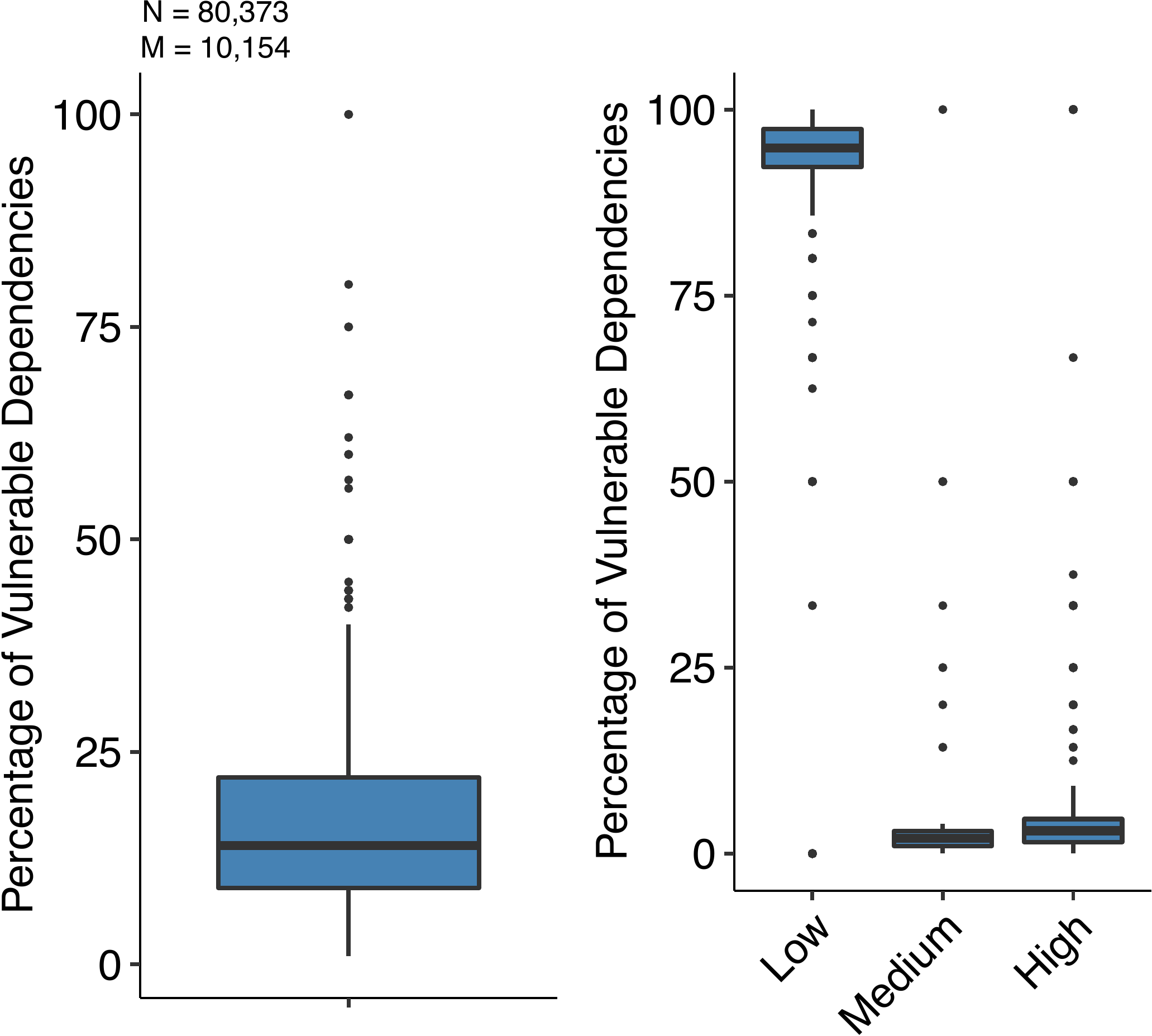}
	\caption{Boxplots showing the distributions of the percentage of overall vulnerable dependencies in the applications (left boxplot), and how these percentages are distributed across threat levels (right boxplot). N and M are the total number of dependencies and the total number of vulnerable dependencies, respectively.
	}
	\label{RQ1Perc}
	
\end{figure}
\noindent \textbf{Results}: 
Of the 6,673 studied applications \textbf{67.93\% (4533 applications) depend on at least one vulnerable dependency}. The affected applications contains a total of 10,154 vulnerable dependencies from 149 distinct vulnerable packages. The 149 packages comprises 23.21\% of the overall vulnerable packages in the npm advisories dataset.

Figure~\ref{RQ1Perc} shows the percentage of vulnerable dependencies per application (left boxplot), and the distribution of vulnerable dependencies at different threat levels (right boxplot). It shows that, on median, 14.29\% of the dependencies in the affected application (i.e., applications with at least 1 vulnerable dependency) are vulnerable. Also, Figure~\ref{RQ1Perc} shows that such percentage of vulnerable dependencies (i.e., 14.29\%) is distributed as follows: \textbf{94.91\% of the vulnerable dependencies are classified as low threat vulnerabilities}, 2.06\% of them are classified as medium, and 3.03\% are classified as high.

\vspace{0.08in}
\begin{table}[h!]
	\centering
	\caption{Mann-Whitney Test (p-value) and Cliff's Delta (d) for the different threat levels.}
	\label{stattest}
	\begin{tabular}{l|r|r}
		\toprule
		\textbf{Threat Levels} & \textbf{\textit{p}-value} &\textbf{ Cliff's Delta \textit{{(d)}}} \\ \midrule
		Low vs. Medium & 2.2e-16 & 0.984 (large) \\
		Low vs. High & 2.2e-16 & 0.970 (large) \\
		Medium vs. High & 2.2e-16 & 0.335 (medium) \\ \bottomrule
	\end{tabular}%
\end{table}
\vspace{0.08in}
To statistically verify our observation, we perform a one-sided non-parametric Mann-Whitney U test~\cite{mcknight2010mann} by comparing the distributions between the different threat levels. 
Table~\ref{stattest} shows the p-values and effect size values. We observe a statistically significant differences between (low and medium), (low and high), (medium and high), at p-value $<$ 0.05 for all comparisons. Furthermore, we observe, using Cliff's delta~\cite{cliff1993dominance}, a large effect size for the differences between low and medium, low and high. Also, we found a medium effect size for the difference between medium and high. This indicates that the differences between the different threat levels are statistically significant.\\

\begin{mdframed}[roundcorner=5pt,linewidth=0.5mm,
	linecolor=black]
	\lipsum[0]
	\textbf{\textit{Our findings show that 67.93\% of the examined applications depend on at least one vulnerable package. However, the vast majority (94.91\%) of these dependencies have low threat.}
}
\end{mdframed}

\vspace{0.2in}
\subsection*{\rqtwo}
{\textbf{\\Motivation}}: Thus far, we have analyzed the vulnerability threats of a single snapshot of each application in our dataset. However, our findings may differ as the applications evolve. For example, a vulnerability with high threat on a given day could have had low threat the week before.


\begin{figure*}[tb!]
	\centering
		\setlength{\abovecaptionskip}{10pt}
	\includegraphics[width=1\linewidth]
	{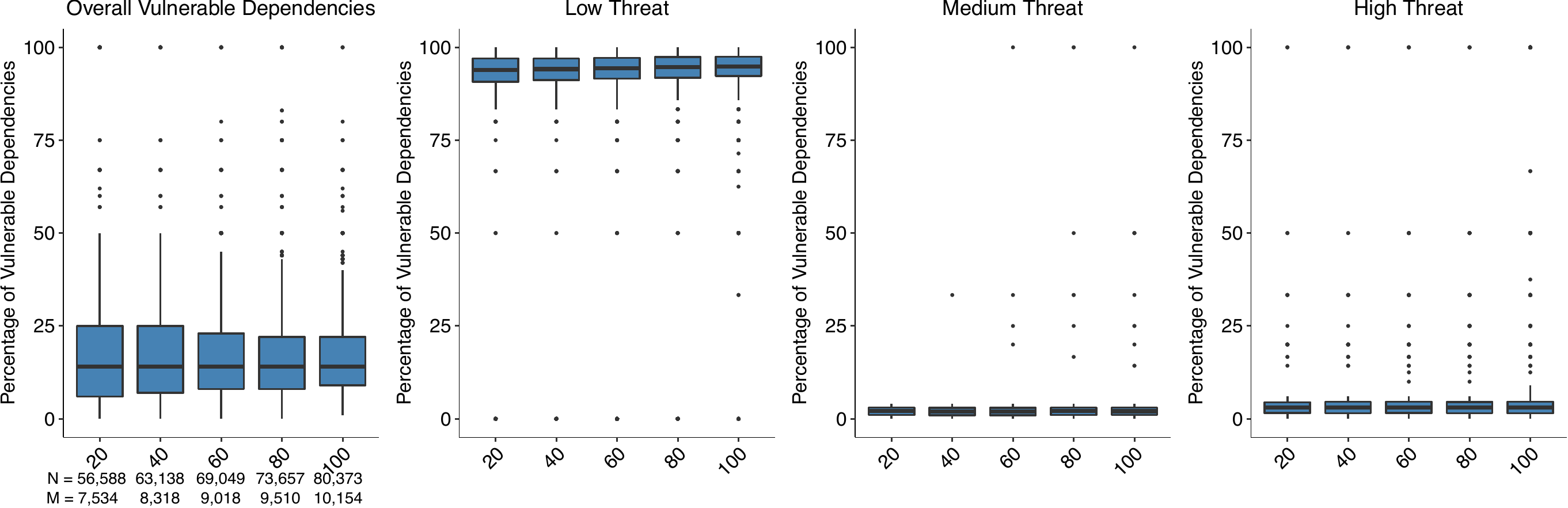}
	\caption{Boxplots showing the percentage of overall vulnerable dependencies and their distribution in each threat level over the studied snapshots. N and M are the total number of dependencies and the total number of vulnerable dependencies, respectively.}
	\label{RQ2Evol}
\end{figure*}


Hence, in this RQ we would like to determine whether our results generalize to different historical snapshots in the application's development lifetime. Such an evolutionary examination allows us to discover whether the trend of the threat levels changes across different stages of an application's lifetime.
\\
\noindent
{\textbf{Approach}}:
Since the different applications are of different lifespans, we want to find a measure that makes comparing them feasible. To do so, we use the number of commits as a way to divide the applications into different intervals. Since commit frequency and time between commits vary from one application to another, we normalize the applications by segmenting the lifetimes of each application into five equal intervals (each containing 20\% of an application's lifetime by {time in days}),
take one snapshot at each interval, then analyze it. Although this might seem like a straightforward task, it poses some challenges, since we have a large applications dataset and the package.json file in them is updated significantly over the application's lifetime.
For this analysis, we only consider the affected applications identified in RQ$_1$.
The last snapshot (at 100\%) is the same snapshot that we analyzed in RQ$_1$ (i.e.,  May 2016).
%

\begin{table}[tbh]
	\centering
	\caption{The percentage of vulnerable applications at different historical snapshots.}
	\label{DisT1}
		\begin{tabular}{l|r}
			\toprule
		\multirow{2}{*}{\textbf{Snapshot}} & \textbf{Vulnerable}  \\ 
										&	\textbf{Applications} \\
			\midrule
			\textbf{20\%} &{ 55.31\%}  \\
			\textbf{40\%} & 58.17\%  \\
			\textbf{60\%} &  60.87\%  \\
			\textbf{80\%} &  63.03\%  \\
			\textbf{100\%} & {67.93\%} \\ \bottomrule
		\end{tabular}
\end{table}
\noindent
{\textbf{Results}}:
Table~\ref{DisT1} shows the percentage of applications that have at least one vulnerable dependency for the 5 analyzed snapshots across their lifetime. We observe that \textbf{the percentage of vulnerable applications steadily increases each snapshot} and varies between 55.31 - 67.93\% in the studied applications.

Figure~\ref{RQ2Evol} illustrates the distributions of the percentage of vulnerable dependencies at each threat level over the studied snapshots. The total number of dependencies (N) and the total number of vulnerable dependencies (M) in the studied applications are shown at the bottom of Figure~\ref{RQ2Evol}. The raw numbers of the dependencies show that the total number of dependencies increases over time, and so does the raw number of vulnerable dependencies.

From Figure~\ref{RQ2Evol}, we observe that the affected applications depend on vulnerable dependencies at an earlier stage (i.e., at 20\%) of their lifetime. However, we also observe that the trend observed in RQ1 remains the same, i.e., the overall percentage of vulnerable dependencies ranges between 14.29\% - 14.68\%. Also, the majority of the vulnerabilities have a low threat level, followed by high and medium threat. To sum up, our analysis shows that all trends observed in RQ$_1$ 
also hold at different stages of the applications, albeit the raw number of dependencies does increase.\\

\begin{mdframed}[roundcorner=5pt,linewidth=0.5mm,
	linecolor=black]
	\lipsum[0]
	\textbf{\textit{As applications evolve, the overall number of vulnerable dependencies is increasing, however, the median percentage of vulnerable dependencies remains mostly constant. Moreover, the majority of vulnerabilities they face remain as low threat vulnerabilities, as these applications evolve.}}
\end{mdframed}

\vspace{-0.2in}
\subsection*{\rqthree}
\textbf{Motivation}:
In the previous research questions, we found that the majority of affected dependencies are impacted by low threat vulnerabilities, throughout applications development history.
However, a sizeable number of projects depend on  
high threat dependencies, which are the most important. This means that those applications depend on vulnerable versions of dependencies even after the vulnerability reports have been discovered (reported)-and-published.
In such cases, the \emph{developers of the applications could know} about the presence of the vulnerability in the dependency, and hence, {should} avoid using that vulnerable version, if a fix is available.
Specifically, we want to know who is to blame - the package maintainers for not providing a version that fixes a known vulnerability - or the application maintainers for not keeping their applications up-to-date. 
Answering this will help us pinpoint the causes for high threat vulnerabilities in npm applications and develop further strategies to solve this problem. 


\noindent
\textbf{Approach}:
%
To perform our investigation and answer who is responsible for the high threat vulnerabilities in applications, we use the same method to determine high threat vulnerabilities as presented in the first two RQs.

%

For each high threat vulnerable dependency, we check the availability of a safe version of the package for the vulnerability at the analyzed snapshot time. Depending on such availability our analysis has one of two outcomes:
\begin{itemize}
	
	\item \textbf{Package-to-blame:} if at the analyzed snapshot, no safe version has been provided by the package maintainers for a publicly known vulnerability. As the publication of a vulnerability comes after a period of 45 days, we consider the package maintainers the responsible for the high threat vulnerability in applications.
	
	\item \textbf{Application-to-blame:} if there is already a released safe version of the vulnerable package but the application continues to rely on an (old) version with a publicly known vulnerability.
	Application developers should monitor their dependencies and update to releases without known vulnerabilities, hence, we consider the application maintainers as responsible for the high threat vulnerability.

\end{itemize}



\begin{table}[tb!]
	\centering
	\caption{The percentage of vulnerabilities caused by the lack of available fix patch (Package-to-blame) vs caused by the lack of dependencies update (Application-to-blame), over the applications snapshot.}
	\label{RQ2T1}
		\begin{tabular}{l|r|r}
			\toprule
			\textbf{Snapshot} &
			\textbf{Package-to-blame} &  \textbf{Application-to-blame} \\
			\midrule
			\textbf{20\%} & {12.06\%} & {87.94\%}  
			\\
			\textbf{40\%}  &  {9.52\%} & {90.48\%} \\
			\textbf{60\%}  & 11.91\% & {88.09\%} \\
			\textbf{80\%}  &  {12.43\%} &{87.57\%} \\
			\textbf{100\%}  & \textbf{9.24\%} & \textbf{90.76\%} \\
			\bottomrule			
		\end{tabular}%
\end{table}

\noindent
\textbf{Results}:
Table~\ref{RQ2T1} shows the percentage of high threat vulnerabilities based on our responsibility analysis. From Table~\ref{RQ2T1}, we observe that \textbf{for high threat vulnerabilities, the application is to blame in 90.76\% of the cases} at the last snapshot (i.e., 100\%).
That means that in 9 out of 10 cases the high threat vulnerability had an available fix, but the applications did not update their dependencies to receive the last fix patch.
Note that this observation holds over all snapshots, with percentages of application-to-blame cases varying from 87.94\% to 90.76\%.

Therefore, and perhaps counter-intuitively, high threat vulnerabilities do not exist because packages have unfixed vulnerabilities, rather the real cause is the fact that these applications fail to keep up or at least to inform themselves well enough about a given dependency version. Hence, a major implication of our study is that application developers need to take updates pushed from their dependencies seriously, or at least actively track their dependencies, since those can lead to very serious effects.


It is important to note that we do not argue about the severity of the vulnerabilities, but rather their likelihood threat of being exploited. Hence, a low severity vulnerability can be very dangerous if everyone knows how to exploit it (high threat level according to our classification). The inverse is also true in that a high severity vulnerability can have a very low chance of being exploited if no one knows about its existence (low threat level).\\
\begin{mdframed}[roundcorner=5pt,linewidth=0.5mm,
	linecolor=black]
	\sloppy
	\lipsum[0]
		\textbf{\textit{{Our findings show that applications not updating their dependencies, are the main cause of high threat (more than 87\%) vulnerabilities.}}
	}
\end{mdframed}
	
	\section{Discussion}
	\label{sec:discussion}
	In this section, we first address the cost of migrating dependencies
for a safer version (Section~\ref{sec:cost}). Then, we discuss how our findings
of vulnerable dependencies may lead to implications to researchers
and practitioners (Section~\ref{sec:imp}).
\vspace{-0.05in}
\subsection{Security Migration Cost}
\label{sec:cost}
Developers of Node.js applications may use dynamic versioning
constraints if they want to install the latest version of a dependency,
allowing them to get the latest updates for security fixes of the package. In fact, npm adopts a semantic version scheme~\cite{semvernp91}, where
package maintainers are encouraged to specify the extent of their
updates in three different levels: 1) patch release, which indicates backward compatible bug fixes, 2) minor release, which indicates backwards
compatible new features and 3) major release, which informs developers of
backwards incompatible changes in the package release. While our
study (RQ$_3$) showed that 90.76\% of high-threat vulnerabilities have a safe
version available for application maintainers (at the snapshot 100\%),
we manually inspected the fixed versions and the applications version constraints and found that in 43.07\% of the cases, the fix is only
available in another major release. For instance, an application
depends on P:1.0.0, and the fix patch was only released for a major
version 2.0.0 and onwards. Hence, to benefit from a fix patch in such a case, developers are required to upgrade their dependencies at the risk of breaking their own code, since a new major release has breaking
changes compared to the version the application depends on. This
imposes significant migration costs, especially for large projects
that depend on dozens of packages. Furthermore, this shows that
using dynamic versioning at the level of patch and minor releases
(as recommended by npm) does not completely prevent high threat
level vulnerabilities for affecting Node.js applications.

\vspace{0.1in}
\subsection{Implications}
\label{sec:imp}
\textbf{Implications to researchers.}
Several studies have addressed the problem of vulnerabilities in software libraries~\cite{decan2018impact,zapata2018towards}. 
Our study, however, complements previous studies by analyzing the risks of vulnerable dependencies in the Node.js applications, aggregating the vulnerability lifecycle through the threat level metric.

Researchers can use our empirical evidence to better understand the risks Node.js applications face due to their high reliance on dependencies. 
Our results show that most vulnerable dependencies found in a application snapshot have a low risk of being exploited when considering the lifecycle of vulnerabilities and how applications update their dependencies. Our results also show that the time element is crucial to understanding the threat of vulnerable dependencies in applications. 
 
Hence, a major implication of our study for researchers is that not all vulnerabilities are equal, and should not be treated and analyzed as such. 
Research needs to account for more than the existence of  vulnerabilities to draw more meaningful analyses regarding software security, particularly for applications in software ecosystems where the level of dependency continues to increase. Research can use our threat-level approach to provide a more refined picture when reporting the impact of vulnerabilities. Researchers can also reuse our approach to help them identify and classify vulnerable dependencies in the applications (in Section~\ref{subsec:scanning}).
%
%

Furthermore, more studies across ecosystems are necessary to get a broader perspective on the threat level of vulnerability  dependencies.
npm is one of the largest ecosystems and since applications depend on an increasingly high number of packages~\cite{decan2018empirical}, Node.js applications may be subjected to higher risk of vulnerable dependencies.
Further investigation could unveil if this pattern holds in other ecosystems.
\\

\noindent
\textbf{Implications to practitioners.}
Our results revealed important takeaways for software practitioners.
First, vulnerable dependencies are common, 67.93\% of the studied Node.js applications had at least one vulnerable dependency at the last studied snapshot.
Practitioners need to be in constant alert to update their dependencies and tools that increase awareness of vulnerabilities, such as Dependabot~\cite{Dependab26:online} and \texttt{npm audit}~\cite{npmaudit59:online} are evermore crucial for the safety of software applications, especially because they warn developers as soon as the vulnerability becomes of a high threat level.

Second, practitioners also need to account for the threat level of a vulnerability to have a more correct understanding of software vulnerabilities in software ecosystems. Our method of analysis can also be used by developers to identify packages that more often raise the threat level in their applications.
Also, while vulnerabilities are widespread in open-source packages in the npm, in most cases package maintainers issue a fix patch for their vulnerability as soon as it becomes public, which is crucial to mitigate the chances of having a vulnerability exploited and cause potential harm to end-users and application maintainers.

Third, our study showed that developers are in need of more tools that go beyond simply warning them of a published vulnerability. For example, they need tools to help them understand: 
1) the costs of migrating to a safer version and whether it is possible to fix a vulnerability without breaking their code,
2) the frequency in which certain dependencies have become vulnerable in the past, in order to grab the threats of depending on such packages and better plan their project maintenance,
3) history of all vulnerable dependencies of their application in order to understand the frequency and the duration in which their application became at the risk of a high threat vulnerability in the past. 
Packages that do not update their code to address reported vulnerabilities incur in a high risk for applications that use them and should be avoided by critical applications.

	\section{Related Work}
	\label{sec:related_work}
	The work most related to our study falls into two main categories - studies on software ecosystems and studies on security vulnerabilities in packages. In the following, we discuss the related work and reflect on how the work compares with ours.

\subsection{Software Ecosystems}
A plethora of recent work focused on software ecosystems. Several works compare different ecosystems. For example, Decan et al.~\cite {decan2018empirical} empirically compared the evolution of 7 popular package ecosystem using different aspects, e.g., growth, changeability, resuability, and fragility. They observed that the number of packages in those ecosystems is growing over time, showing their increasing importance. 

Other work focused specifically on npm~\cite{fard2017javascript,kula2017impact, wittern2016look}. For example, Fard et al.~\cite{fard2017javascript} examined the evolution of dependencies within an npm project, and showed that there is a heavily interdependence, with the average number of dependencies being 6 and growing over time. Wittern et al.~\cite{wittern2016look} investigated the evolution of npm using metrics such as dependencies between packages, download count, and usage count in JavaScript applications. They found that packages in the npm ecosystem are steadily growing. Such amounts of packages make the spread and discovery of vulnerabilities much worse, given the heavy dependence on such packages and the potential security problems in those packages.

Other studies pointed out the fragility of software ecosystems and provided insights on the challenges application developers face. For example, Bogart et al.~\cite{bogart2015breaks,bogart2016break} examined the Eclipse, CRAN, and npm ecosystems, focusing on what practices cause API breakages. They found that a main reason for breaking changes are the updates of a dependency. This finding may explain why application developers are hesitant to update and explain why we see high threat vulnerabilities impacting applications that do not update in time. 
%
%
%

Our study differs from the prior work since we focus on the threat level of dependency vulnerabilities in Node.js applications. Moreover, we examine how this threat level changes as applications evolve and examine the reason that high threat dependency vulnerabilities exist. That said, much of the aforementioned work motivated us to study npm and focus on examining vulnerabilities in application dependencies.
\vspace{-0.05in}
\subsection{Security Vulnerabilities in Dependencies/Packages}
%

Several works in the literature studied vulnerabilities that come from dependencies~\cite{di2009life,pham2010detection,cox2015measuring,massacci2011after,derr2017keep}. For example, Di Penta et al.~\cite{di2009life} and Pham et al.~\cite{pham2010detection} conducted empirical studies to analyze the evolution of vulnerabilities in source code, and found that most vulnerabilities are recurring due to software code reuse or libraries (i.e., dependencies). Cox et al.~\cite{cox2015measuring} evaluated ``dependency freshness'' to understand the relationship between outdated dependencies and vulnerabilities using industry benchmarks, and found that vulnerabilities were four times as much likely to have existed in outdated systems than in updated systems. Relative studies by Massacci et al.~\cite{massacci2011after} and Derr et al.~\cite{derr2017keep} are in line with~\cite{cox2015measuring}. In general, they both reported that vulnerabilities appeared commonly in non-maintained code and old versions, and this could be fixed by just an update to a newer version. Our study complements these studies by examining the threat of these vulnerabilities in the dependent applications.

More specifically, vulnerabilities that affect packages in ecosystems have been studied broadly~\cite{kula2018developers,pashchenko2018vulnerable}. For example, Kula et al.~\cite{kula2018developers} analyzed the Maven ecosystem on more than 4,000 GitHub projects that correspond to 850,000 library migrations, and found that projects were heavily dependent on these libraries, and most projects (i.e. 81.5\%) had outdated libraries. The study also mentioned (based on interviews conducted with developers) that developers do not update dependencies, and 69\% of the interviewed developers tend to be not aware of their vulnerable dependencies. Pashchenko et al.~\cite{pashchenko2018vulnerable} studied the vulnerability impact of 200 open-source Java libraries commonly used in SAP~\cite{SAPSoftw78} organisation, and found that 20\% of the vulnerable dependencies are not deployed, and hence, they are not exploitable in practice. Moreover, they found that the majority of the vulnerable dependencies (81\%) can be fixed by a simple upgrade to a newer safe version, suggesting that software development companies have to  allocate their audit tools correctly.

Other recent work focused on analyzing vulnerabilities in the npm ecosystem. For example, Hejderup’s~\cite{hejderup2015dependencies}
analysed only 19 vulnerable packages
and found that the number of vulnerabilities in them is growing over time. Similarly, Decan et al.~\cite{decan2018impact} analyzed the vulnerabilities in the npm ecosystem and found that the number of vulnerabilities is growing over time. Also, they reported that it takes a long time to discover vulnerabilities that affect npm packages. Our study complements this study by analyzing the risks of vulnerable dependencies in the Node.js applications (not addressed by the study~\cite{decan2018impact}), aggregating the vulnerability lifecycle through the threat level metric. 
A recent study by Zapata et al.~\cite{zapata2018towards} assessed the danger of having vulnerabilities in dependent libraries by analyzing function calls of the vulnerable functions.
They manually analyzed 60 projects that depend on vulnerabilities, and found that 73.3\% of them were actually safe because they did not make use of the vulnerable functionality of their dependencies, showing that there is a considerable overestimation on previous reports. Our study identifies yet another source of overestimation by including a time-based analysis into a large and comprehensive set of \textit{applications} (i.e., 6,673 Node.js applications).
Zimmermann et al.~\cite{zimmermann2019small} studied the security threat of the npm ecosystem dependencies by mainly analysing the maintainers role and responsibilities for vulnerable packages. They mainly observed that a very small number of maintainers' accounts (i.e., 20 accounts) could be used to inject malicious code into thousands of npm packages, a problem that has been increasing over time. Zerouali et al.~\cite{zerouali2019impact} studied npm vulnerable packages in Docker containers, and found that they are common in the containers, suggesting that Docker containers should keep their npm dependencies updated.

To assess the impact of vulnerable dependencies in the dependent Java applications, Plate et al.~\cite{plate2015impact} proposed an approach that provides a fine-grained assessment of the vulnerabilities that affect dependencies in dependent Java applications. In particular, the approach first determines whether or not the application makes use of the library that is known to be vulnerable. Then, the approach tries to determine whether or not the application executes the fragment of the dependency where the vulnerable code is located. Furthermore, Ponta et al.~\cite{ponta2018beyond} built upon their previous approach in~\cite{plate2015impact} to generalize their vulnerability detection approach by using static and dynamic analysis to determine whether the vulnerable code in the library is reachable through the application call paths. Their proposed approach is implemented in a tool called, Vulas, which is an official software used by SAP to scan its Java code.

Our study focuses on analyzing the threat of npm vulnerabilities in dependencies, which affected applications that rely on them. In many ways, our study complements the related work since, (1) instead of studying security vulnerabilities that exist in packages, we particularly focus on the threat of such vulnerable packages by real-world open source applications; (2) we provide a threat classification for software vulnerabilities based on their lifetime, and we use our classification and perform an empirical study on Node.js applications.

	\section{Threats to Validity}
	\label{sec:threats}
\textbf{Construct Validity} considers the relationship between theory and observation, in case the
measured variables do not measure the actual factors. Our dataset contains 654
vulnerabilities available in the npm advisories dataset. There might be other vulnerable packages that have been discovered but not yet reported. However, we leveraged up-to-date dataset from npm advisories, which we believe contains complete information about the vulnerable packages reported to them.

With respect to the affected dependencies, we only take into consideration production dependencies (i.e., dependencies that are required to install and run the application). We ignore other types (e.g., development dependencies), because they have no direct impact on the production environment.

This paper only considered direct dependencies. Our results may vary if indirect dependencies are considered, however, due to computation requirements, we focused on the direct dependencies of applications. In the future, we plan to expand our technique to consider indirect dependencies when considering the threat of vulnerabilities.

We did not consider whether the vulnerable functionality in the package actually affects the application, i.e., whether the applications uses the vulnerable code of the package. Considering this would be challenging, since our dataset is composed of thousands of applications. That said, our analysis is in line with prior work in the area of software ecosystems, which also examine dependencies in the package.json file to associate packages to applications.
\\

\noindent \textbf{External Validity} is related to the generalizability of our findings. Our study is based on Node.js applications that use npm. Hence our results may not generalize to applications written in other languages. However, the key concepts and design of our study can be applied on other package dependency networks. Although npm is a single case, examples from the past have shown that individual cases contribute to the building of a general empirical evidence software engineering~\cite{flyvbjerg2006five}. 

Our dataset contains 6,673 JavaScript applications that use npm packages. Our dataset might be considered small when it is compared to the whole population of JavaScript applications. However, our dataset is of high quality, since we filtered out applications that are immature and have less development history, by using the filtering criteria used by Kalliamvakou et al.~\cite{kalliamvakou2014promises}.

	\section{Conclusion and Future Work} 
	\label{sec:conclusion}
	Our study examined software vulnerabilities in npm dependencies with respect to their threat on the dependent Node.js applications. First, we defined three levels of threat for software vulnerabilities in dependencies based on their lifecycle and performed an empirical study on 6,673 Node.js applications to assess how threatening the vulnerable dependencies that exist in these applications really are. Our findings indicate that the vast majority of vulnerable dependencies have low threat on applications that depend upon them. 
Although 67.93\% of the examined applications depend on at least one vulnerable package, 94.91\% of the vulnerable dependencies are classified as having low threat.
Moreover, we examined why these applications end up depending on high threat vulnerable versions of these dependencies. We observed that, in the case of high threat vulnerabilities, the applications are to blame in more than 87\% of the cases, i.e., a fix for the vulnerable dependency is available but not patched in the application.
These findings show that the assumption that all vulnerabilities that exist in packages will impact applications the same way is not correct and that vulnerable packages are not always to blame. Finally, our further analysis shows that all of the observed trends hold across the different stages of the applications' lifetime.

In the future, we plan to further elaborate on the impact of the various threat vulnerabilities on the applications' functionality level. Other data sources  can be added to enhance the risk assessment, e.g., severity, exploitability, etc.  We also aim to examine if our findings hold for applications written in different programming languages.

\bibliographystyle{IEEEtran}
\balance
\bibliography{bibliography}
	\ifCLASSOPTIONcaptionsoff
	\newpage
	\fi

	\newpage
	\begin{IEEEbiography}[{\includegraphics[width=1in,height=1.25in,clip,keepaspectratio]{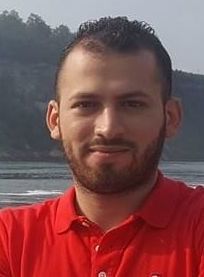}}]{Mahmoud Alfadel} 
	\par is a PhD student in the Department of Computer Science and Software Engineering at Concordia University in Canada. Mahmoud obtained
	a B.Sc. in Informatics from Damascus University in
	Syria (2014) and an M.Sc. in Computer Science from King
	Fahd University of Petroleum and Minerals in Saudi Arabia
	(2017). His research interests include empirical software engineering, mining software repositories, and vulnerabilities in software ecosystems.
	\par 
	\end{IEEEbiography}

	\begin{IEEEbiography}[{\includegraphics[width=1in,height=1.25in,clip,keepaspectratio]{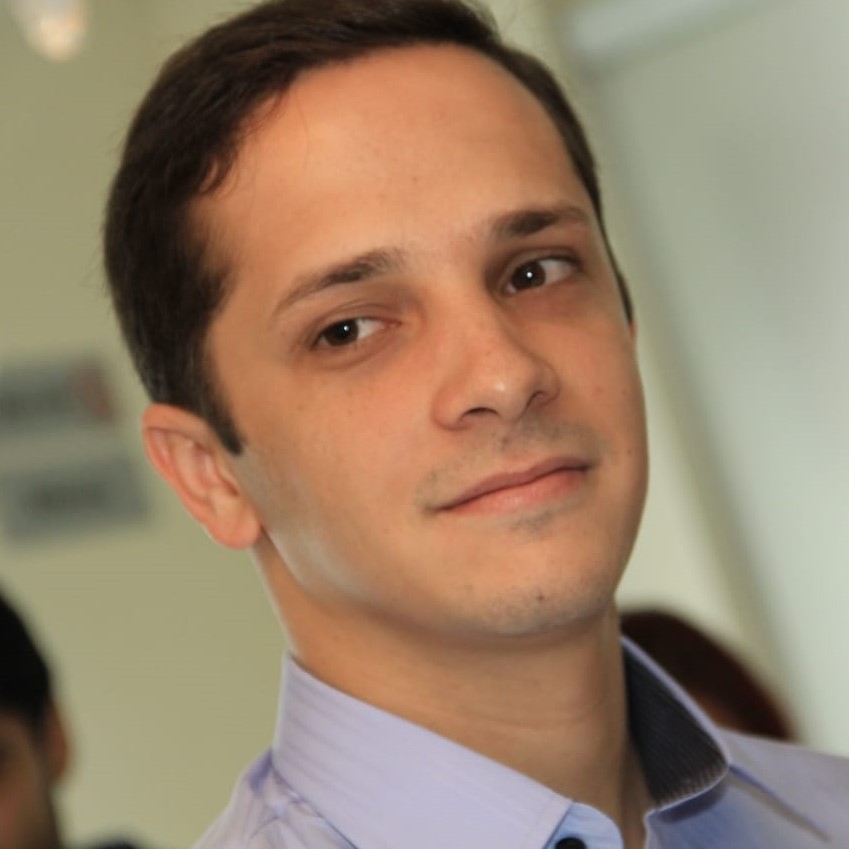}}]{Diego Elias Costa}
		\par is a postdoctoral researcher in the Department of Computer Science and Software Engineering at Concordia University. 
		He received his PhD in Computer Science from Heidelberg University.
		His research interests cover a wide range of software engineering and performance engineering related topics, including mining software repositories, empirical software engineering, performance testing, memory-leak detection, and adaptive data structures. 
		You can find more about him at \url{http://das.encs.concordia.ca/members/diego-costa/}.
		\par
	\end{IEEEbiography}

	\begin{IEEEbiography}[{\includegraphics[width=1in,height=1.25in,clip,keepaspectratio]{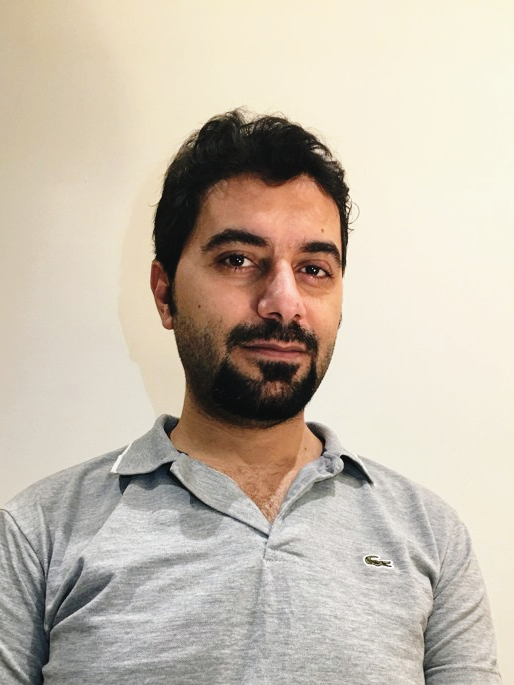}}]{Mouafak Mkhallalati} 
		\par is a Masters graduate. Mouafak obtained his M.Sc. at Concordia University in Canada. Mouafak has hands-on experience as a software engineer working on a variety of technologies, languages, and paradigms. His research interests include Software Security, Software Testing, and mining software repositories.
		\par 
	\end{IEEEbiography}

	\begin{IEEEbiography}[{\includegraphics[width=1in,height=1.25in,clip,keepaspectratio]{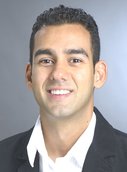}}]{Emad Shihab} 
		\par is an associate professor in the
		Department of Computer Science and Software Engineering at Concordia University. He received his PhD from Queens University. Dr. Shihab's research interests are in Software Quality
		Assurance, Mining Software Repositories, Technical Debt, Mobile Applications and Software Architecture. He worked as a software research intern at Research In Motion in Waterloo, Ontario
		and Microsoft Research in Redmond, Washington. Dr. Shihab is a member of the IEEE and ACM. More information can be found at \url{http://das.encs.concordia.ca}.
		\par
	\end{IEEEbiography}

	\begin{IEEEbiography}[{\includegraphics[width=1in,height=1.25in,clip,keepaspectratio]{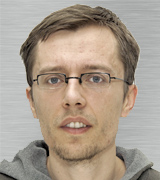}}]{Bram Adams} 
		\par is an associate professor at Polytechnique Montreal, where he heads the Lab on Maintenance, Construction, and Intelligence of
		Software. His research interests include release
		engineering in general, as well as software integration, software build systems, and infrastructure as code. Adams obtained his PhD in computer science engineering from Ghent University. He is a steering committee member of the
		International Workshop on Release Engineering
		(RELENG) and program co-chair of SCAM 2013,
		SANER 2015, ICSME 2016 and MSR 2019. More can be found at \url{http://mcis.polymtl.ca/bram.html}.
		\par 
	\end{IEEEbiography}
    
	
	

\end{document}